\documentclass[aps,prb,twocolumn,notitlepage,superscriptaddress,showpacs]{revtex4-1}
\usepackage{amsmath,latexsym,amssymb,bm}
\usepackage{color}
\usepackage{graphicx,subfigure}
\usepackage{multirow}
\usepackage{natbib}

\begin{document}

\title{Single-hole wave function in two dimensions: A case study of the doped
Mott insulator}
\author{Shuai Chen}
\affiliation{Institute for Advanced Study, Tsinghua University, Beijing,
100084, China}
\author{Qing-Rui Wang}
\affiliation{Department of Physics, The Chinese University of Hong Kong, Shatin, New Territories, Hong Kong, China}

\author{Yang Qi}
\affiliation{Center for Field Theory and Particle Physics, Department of Physics, Fudan University, Shanghai 200433, China}
\affiliation{State Key Laboratory of Surface Physics, Fudan University, Shanghai 200433, China}
\affiliation{Collaborative Innovation Center of Advanced Microstructures, Nanjing 210093, China}

\author{D. N. Sheng}
\affiliation{Department of Physics and Astronomy, California State University, Northridge, CA, 91330, USA}

\author{Zheng-Yu Weng}
\affiliation{Institute for Advanced Study, Tsinghua University, Beijing,
100084, China}
\affiliation{Collaborative Innovation Center of Quantum
Matter, Tsinghua University, Beijing, 100084, China}

\begin{abstract}
We study a ground-state ansatz for the single-hole doped $t$-$J$ model in
two dimensions via a variational Monte Carlo (VMC) method. Such a
single-hole wave function possesses finite angular momenta generated by
hidden spin currents, which give rise to a novel ground state degeneracy in
agreement with recent exact diagonalization (ED) and density matrix
renormalization group (DMGR) results. We further show that the wave function
can be decomposed into a quasiparticle component and an incoherent momentum
distribution in excellent agreement with the DMRG results up to an $8\times 8
$ lattice. Such a two-component structure indicates the breakdown of
Landau's one-to-one correspondence principle, and in particular, the
quasiparticle spectral weight vanishes by a power law in the large
sample-size limit. By contrast, turning off the phase string induced by the
hole hopping in the so-called $\sigma\cdot t\text{-}J$ model, a conventional
Bloch-wave wave function with a finite quasiparticle spectral weight can be
recovered, also in agreement with the ED and DMRG results. The present study
shows that a singular effect already takes place in the single-hole-doped
Mott insulator, by which the bare hole is turned into a non-Landau
quasiparticle with translational symmetry breaking. Generalizations to
pairing and finite doping are  briefly discussed. 
\end{abstract}

\date{\today}
\maketitle
\tableofcontents

\affiliation{Institute for Advanced Study, Tsinghua University, Beijing
100084, China}

\affiliation{Department of Physics, The Chinese University of Hong Kong,
Shatin, New Territories, Hong Kong, China}

\affiliation{Department of Physics and Astronomy, California State
University, Northridge, CA, 91330, USA}

\affiliation{Institute for Advanced Study, Tsinghua University, Beijing
100084, China} \affiliation{Collaborative Innovation Center of Quantum
Matter, Tsinghua University, Beijing 100084, China}

\section{Introduction}

High-temperature superconductivity (HTS) in the cuprate \cite%
{Bednorz1986Possible} is widely considered to be a strong correlation effect
of the doped Mott insulator\cite{Anderson1987Resonating}, where the pairing
is not due to the phonon mechanism  as in  the original BCS theory \cite%
{Bardeen1957Theory}. In such a pure interacting electron system, the nature
of the ``normal state" prior to superconducting transition is crucial \cite%
{Anderson1987Resonating, Anderson-book, Lee2006Doping} in understanding the
HTS mechanism.

In a conventional normal metal (Fermi liquid), each new particle injected
into the system should behave like a Landau's quasiparticle at low energies.
By contrast, in a  non-Fermi-like Luttinger liquid\cite{Haldane1994Luttinger}
(LL), vanishing quasiparticle spectral weight has been identified in the
one-dimensional (1D) doped Mott insulators, described by the Hubbard \cite%
{Ogata1990Bethe, Ogata1991phase, Anderson1993Asymptotic} and $t$-$J$ model 
\cite{Weng1991one,Zhu2016Exact}. The generalization of a possible LL state
to two dimensions (2D) in connection with the HTS cuprate has been
conjectured \cite{Anderson-book} early on, but so far it has not been fully
substantiated by  either  theory or experiment.

The key issue is how a doped hole propagates in the 2D quantum spin
background of a doped Mott insulator as compared to a Fermi liquid. To this
end, the study of a single-hole case has been of central interest as the
simplest case of ``normal state''. Considerable efforts have been invested
in studying the single hole's motion in a 2D antiferromagnet described by
the $t$-$J$ model using both analytical and numerical techniques. Despite of
a strong distortion induced by the hole in the spin background, which is
generally known as the spin polaron effect, many early studies have
concluded that the hole would still behave like a coherent quasiparticle
with a finite spectral weight in the long-wavelength, low-energy regime.
Shraiman and Siggia \cite{Shraiman1988mobile, Shraiman1990mobile}
proposed a semiclassical variational wave function and an effective
Hamiltonian which  treats the twisted spin configuration as a dipolar
distortion. In the self-consistent Born approximation \cite{Brinkman1970,
Schmitt-Rink1988Spectral, Kane1989, Martinez1991spin,Liu1991Spectral} (SCBA)
approaches, spin magnon excitations renormalize the effective mass of a hole
to result in a much reduced bandwidth as compared to the bare band
parameters. The corresponding dispersion has the energy minima at momenta $%
\mathbf{K}^0=(\pm\pi/2, \pm\pi/2)$, which agrees with the exact 
diagonalization (ED) finite-size calculations \cite{Dagotto1994correlated, Leung1995dynamical}. Assuming a finite quasiparticle
spectral weight and local minima at $\mathbf{K}^0$, later efforts \cite%
{Leung1997Comparison, Tohyama2000Spin, Lee2003Low, Tohyama2004Asymmetry}
have been further devoted to issues like the detailed dispersion by
including the next-nearest-neighbor hoppings, $t^{\prime }$ and $t^{\prime
\prime }$, in comparison with the angle-resolved photoemission spectroscopy
(ARPES)\cite{Wells1995Eversus,Ronning1998Photoemission,
Armitage2001Anomalous, Damascelli2003angle}.

However, a recent numerical study by ED and density matrix renormalization
group (DMRG) has revealed \cite{Zheng2018} an important fact that,
accompanying the single hole in the ground state, hidden spin currents are
generically present in the background, which has been essentially overlooked
in the previous numerical studies\cite{Dagotto1994correlated, Leung1995dynamical} of the 2D $t$-$J$ model. Namely,
despite of a total momentum at $(\pm\pi/2, \pm\pi/2)$ under a periodic
boundary condition (PBC), the doped hole itself may not always carry the
full momentum since a part of it has been carried away by the spin current
into the neutral spin background \cite{Zheng2018}. Furthermore, under an
open boundary condition (OBC) which maintains the lattice $C_4$ rotational
symmetry, the ground state is characterized by finite angular  momentum  $%
L_z=\pm 1$ (mod 4), which is tied to the chiralities of the background spin
currents with a double degeneracy \cite{Zheng2018}. These are in sharp contrast to the above-mentioned
spin polaron picture obtained by SCBA, which has suggested that the hole
should be dressed by a \emph{rigid} spin distortion at low energies that
only renormalizes its effective mass, which still satisfies the Landau's
one-to-one correspondence hypothesis such that the total momentum is fully
carried by low-lying quasiparticle excitations.

Thus, the novel ground state degeneracy and associated spin currents found
by numerical calculations \cite{Zheng2018} have clearly indicated that a
single-hole ground state of the $t$-$J$ model is not simply described by a
conventional quasiparticle. Here one has to treat the local coupling between the hole and spin
background more carefully as it may be much more singular than previously
believed. This is critically important in order to meaningfully extrapolate
the single-hole results into a large scale or a finite doping regime, where
the local correlation dominated by the single hole's mutual influence with
the spin background must be correctly taken into account as a proper
starting point in order to understand pairing and other long-wavelength
physics.

Therefore, in this work, we shall revisit the single-hole ground state of
the 2D $t$-$J$ model by studying a new ground state ansatz by variational
Monte Carlo (VMC) method. Such a variational state is the one-hole limit of
the ground state ansatz previously constructed in the phase string theory 
\cite{Weng2011superconducting, Weng2011mott}. It has been already applied to
the one-hole cases in the 1D\cite{Wang2014Sign, Zhu2016Exact} as well as
two-leg ladder\cite{Wang2015variational} systems, with the VMC results well
reproducing various singular features observed in the DMRG calculations. As
opposed to the above-mentioned spin polaron \cite{Shraiman1988mobile,
Shraiman1990mobile,Schmitt-Rink1988Spectral, Kane1989,
Martinez1991spin,Liu1991Spectral} or ``spin bag'' picture\cite%
{Schrieffer1988Spin, Weng1988d}, a hole hopping on an antiferromagnetically
correlated spin background will generally create a string-like spin mismatch
defect, known as the phase string \cite{Weng1996Phase, Weng1997Phase,
Wu2008Sign}, which cannot be completely ``self-healed'' by the spin
superexchange dynamics. In the SCBA scheme \cite{Schmitt-Rink1988Spectral,
Kane1989, Martinez1991spin,Liu1991Spectral}, the longitudinal $S^z$-string 
\cite{Bulaevski1968New} induced by hopping is assumed to be
reparable via the spin flip process of the Heisenberg term. However, the
hopping of the hole will simultaneously generate the transverse $S^x$ and $%
S^y$ strings as well. An exact mathematical formulation\cite{Weng1996Phase,
Weng1997Phase, Wu2008Sign} has shown that after the $S^z$-string along the
spin $z$-direction is ``repaired'' through spin flips, the transverse
strings, represented by a sequence of \emph{signs} known as the phase
string, cannot be erased simultaneously and will be generally left in the
hole path, which plays a role like the Berry phase to result in strong
quantum interference once the whole paths of the hole are summed over in the
Feynman's path-integral fashion. Namely, the doped holes will always create
irreparable spin mismatch ``strings'' on its path to singularly influence
its motion on a quantum spin background. Indeed, by precisely turning off
such a phase string effect in the $t$-$J$ model to result in the so-called $%
\sigma\cdot t$-$J$ model \cite{Zhu2013Strong}, the above-outlined spin
current pattern and novel ground state degeneracy all disappear to recover a
true quasiparticle description as shown by ED and DMRG in Ref. %
\onlinecite{Zheng2018}.

Specifically, in this paper, we study such a single-hole variational ground
state ansatz by VMC, in which the bare hole is ``twisted'' by producing a
nonlocal phase shift due to the phase string effect based on the $t$-$J$
model. It can correctly describe a one-hole quantum state with the conserved
hole number $N_h=1$, spin $S=S_z=1/2$, and an angular momentum $L_z=\pm1$
corresponding to a discrete $C_4$ rotation symmetry under the OBC. Namely,
it has a novel double degeneracy for a given $S_z=1/2$, corresponding to two
chiralities of the neutral spin currents, all in agreement with the ED and
DMRG results \cite{Zheng2018}. Both the momentum distribution $n^h(\mathbf{k}%
)$ and quasiparticle weight $Z_\mathbf{k}$ are in excellent agreement with
the DMRG results up to an $8\times 8$ lattice, showing that the hole
wave function in the ground state can be decomposed into a quasiparticle
component and an incoherent component with a broad continuous momentum
distribution. In particular, $Z_\mathbf{k}$ is indeed peaked at four momenta 
$\mathbf{K}^0=(\pm\pi/2,\pm\pi/2)$. But the finite-size scaling of $Z_{%
\mathbf{K}^0}$ shows a power-law decay, indicating the breakdown of the
quasiparticle picture in the thermodynamic limit. The translational symmetry
is explicitly broken in such a variational ground state. However, by turning
off the phase string, all the above novel features disappear and the
variational ground state recovers a simple Bloch-wave state in the $%
\sigma\cdot t$-$J$ model. Finally, a natural generalization of the present
scheme to pairing and finite doping  are  briefly discussed in the end of
the paper.

The rest of the paper is organized as follows. In Sec.~\ref{Sec::model and
result}, we introduce the $t$-$J$ model on a 2D square lattice and outline
the key results. In Sec.~\ref{Sec::Gs ansatz}, we present the detailed
composite structure of the single-hole variational ground state, which
possesses the same quantum numbers as those identified in the previous ED
and DMRG calculations. We further identify the two-component structure of
the single-hole state of the $t$-$J$ model. Two physical quantities, i.e.,
the momentum distribution $n^h(\mathbf{k})$ and quasiparticle spectral
weight $Z_{\mathbf{k}}$, are presented. For the $\sigma\cdot t$-$J$ model, a
conventional Bloch-wave state is recovered. Finally, the summary and
discussion along with some perspectives are given in Sec.~\ref{Sec::Disc}.

\begin{figure}[t]
\begin{center}
\includegraphics[width=0.5\textwidth]{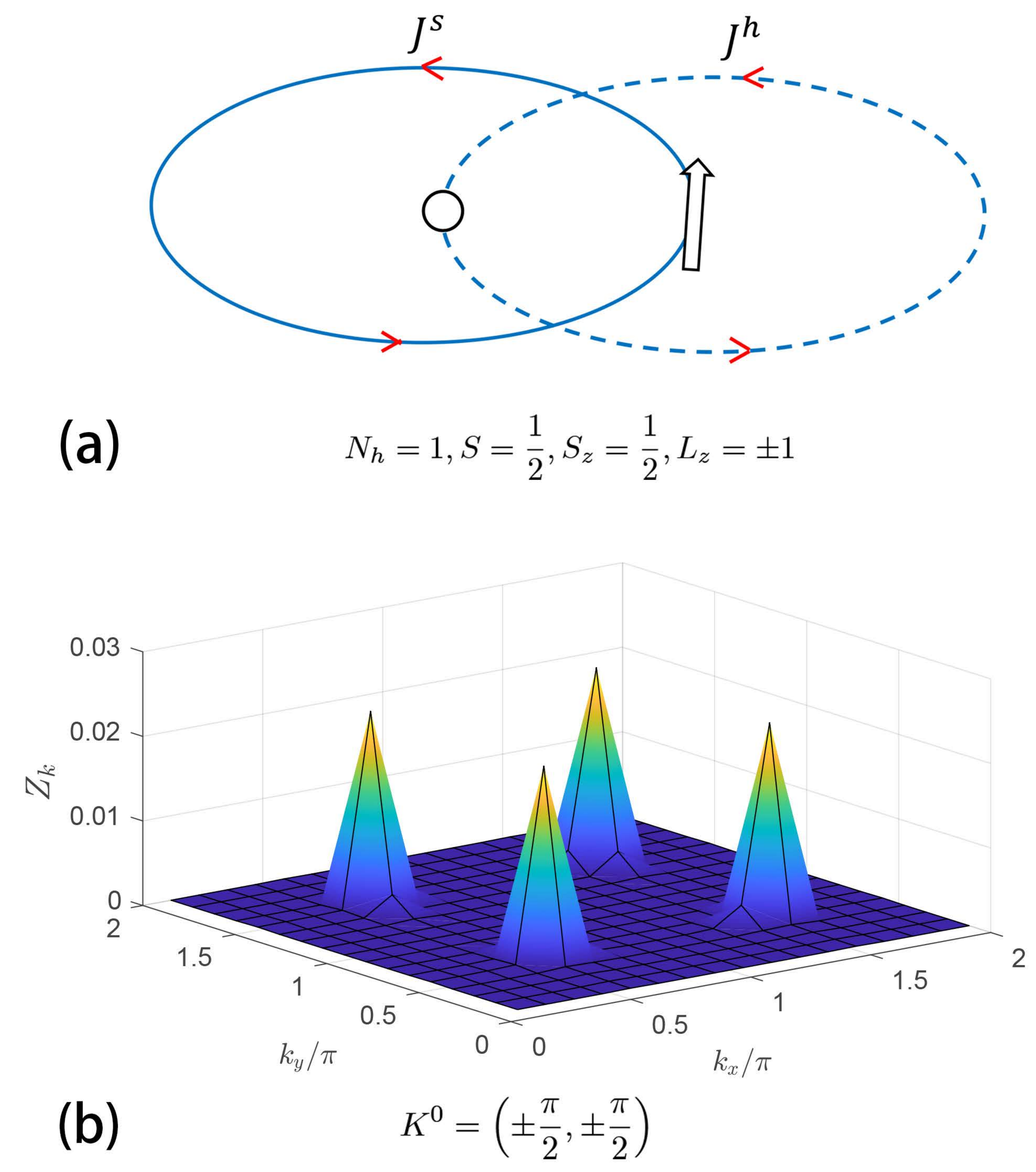}
\end{center}
\par
\renewcommand{\figurename}{Fig.}
\caption{(Color online.) (a) Schematic illustration of the single-hole
wave function ansatz [Eq.~(\protect\ref{gs})], in which the mutual
entanglement between a hole and surrounding spins are explicitly
characterized by the phase string operator [Eqs.~(\protect\ref{tildec}) and (%
\protect\ref{Omega})]. Such a single hole ground state can be labelled by
the following quantum numbers: hole number $N_{\text{h}}=1$, spin $S=1/2$, $%
S_z=\pm 1/2 $, and for a lattice with a discrete $C_4$ rotational symmetry,
an angular momentum $L_z=\pm1$ with nontrivial spin and hole currents, $J_s$
and $J_h$, in agreement with the ED and DMRG results \protect\cite{Zheng2018}%
; (b) The quasiparticle spectral weight of the ground state (\protect\ref{gs}%
) shows four sharp peaks at momenta $(\pm\protect\pi/2, \pm\protect\pi/2)$
at a finite size system ($N=16\times16$). }
\label{Fig1}
\end{figure}

\section{The model and key results}

\label{Sec::model and result}

\subsection{The $t$-$J$ model}

The model we  consider in this work is the standard $t\text{-}J$ model on a
two-dimensional isotropic square lattice with the Hamiltonian $H_{t-J}=%
\mathcal{P}\left( H_{t}+H_{J}^{{}}\right) \mathcal{P}$, in which 
\begin{align}  \label{Hamiltonian}
H_{t} &=-t\sum_{\left\langle ij\right\rangle \sigma }c_{i\sigma }^{\dagger
}c_{j\sigma }^{{}}+\text{H.c}~, \\
H_{J} &=J\sum_{\left\langle ij\right\rangle }\left( \mathbf{S}_{i}\cdot 
\mathbf{S}_{j}-\frac{1}{4}n_{i}n_{j}\right),
\end{align}
where the projective operator $\mathcal{P}$ imposes the no-double-occupancy
constraint on each site. Here, $c_{i\sigma}$ annihilates an electron at site 
$i$ with spin $\sigma$, and $n_{i}^{}=\sum_{\sigma}c_{i\sigma}^\dagger
c_{i\sigma}^{}$ and $\mathbf{S}_{i}$ are the electron number and spin
operator, respectively. We fix $t/J=3$ in making comparison of the present
VMC study of the variational ground state with the ED and DMRG results.

\subsection{Key results}

In this work, we study the following one-hole ground state ansatz for the 2D 
$t$-$J$ model 
\begin{equation}  \label{gs}
|\Psi _{\text{G}}\rangle_{\text{1h}}=\sum_{i}\varphi _{\text{h}}\left(
i\right) \tilde{c}_{i\downarrow }^{}|\text{RVB}\rangle ~,
\end{equation}%
and show that it can systematically reproduce the numerical ED and DMRG
results \cite{Zheng2018} via a VMC calculation. Here $|\text{RVB}\rangle $
denotes a \emph{half-filled} spin background\cite{Liang1988some}, which is
the ground state of the Heisenberg model $H_J$. The doped hole is created by
the annihilation operator 
\begin{equation}  \label{tildec}
\tilde{c}_{i\downarrow }^{}={c}^{}_{i\downarrow }e^{-i\hat{\Omega}_i}
\end{equation}
which removes an electron of spin $\downarrow$ (without loss of generality)
from the ``vacuum'' state $|\text{RVB}\rangle $, and at the same time,
produces a nonlocal many-body ``phase shift'' $\hat{\Omega}_i$ in the spin
background. The latter is defined by \cite{Weng2011superconducting,
Weng2011mott} 
\begin{equation}  \label{Omega}
\hat{\Omega}_i = \sum_{l}\theta _{i}\left( l\right) n_{l}^{\downarrow}~,
\end{equation}
in which in general $\theta _{i}\left( l\right)$ satisfies 
\begin{equation}
\theta _{i}\left( l\right)=\theta _{l}\left( i\right)\pm \pi
\end{equation}
and in particular, it takes the form $\theta _{i}\left( l\right) =\pm \text{%
Im}\ln \left( z_{i}-z_{l}\right) $ in 2D, with $z_{i}=x_{i}+iy_{i}$ as the
complex coordinate of site $i$, and $n_{l}^{\downarrow}$ denotes the number
operator of $\downarrow$ spin at site $l$. Finally, $\varphi _{\text{h}} $
in Eq.~(\ref{gs}) is a variational parameter representing the wave function
of the doped hole, which is to be determined by minimizing the variational
energy in the VMC calculation.

\begin{figure*}[tb]
\begin{center}
\includegraphics[width=\textwidth]{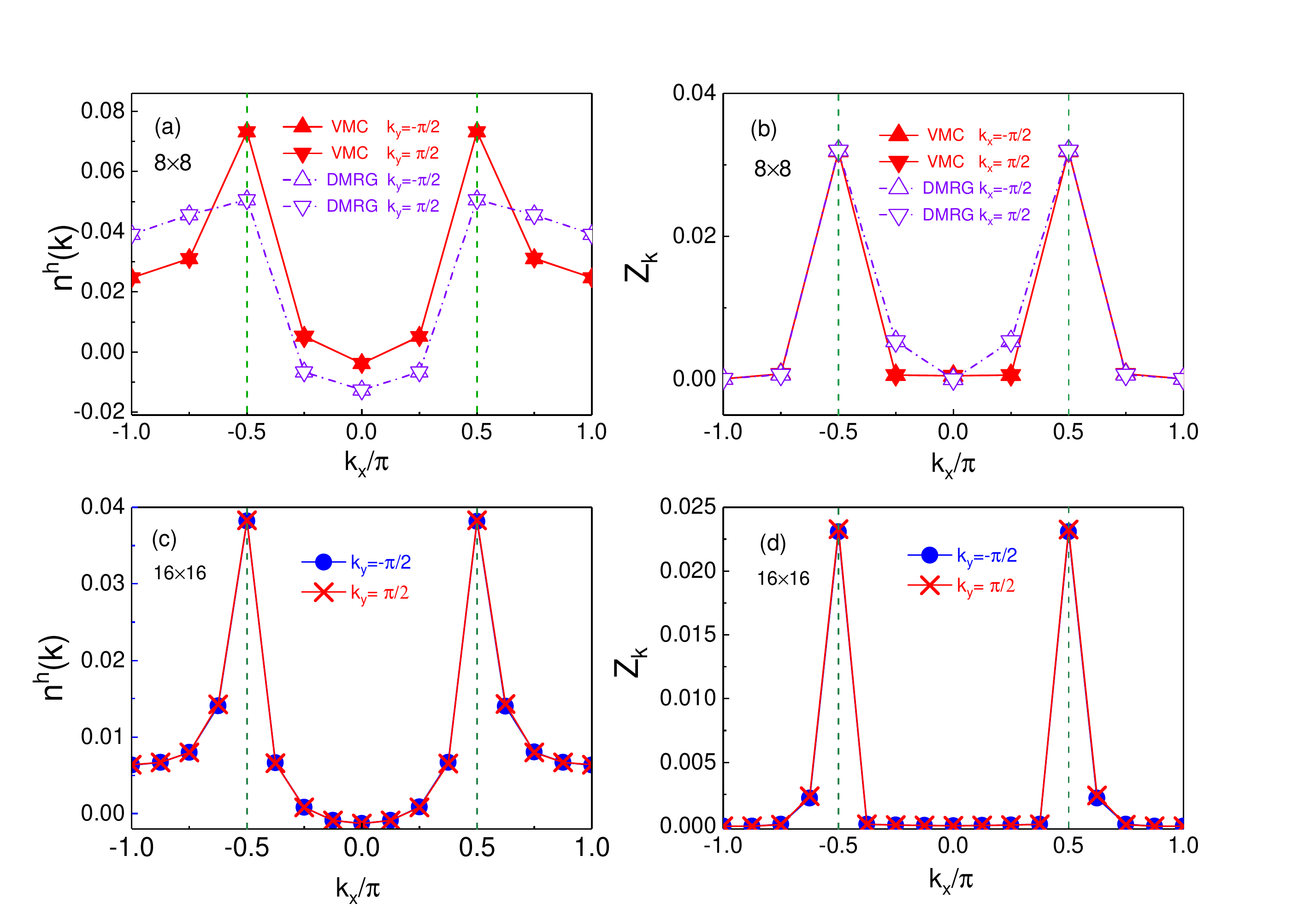}
\end{center}
\par
\renewcommand{\figurename}{Fig.}
\caption{(Color Online.) Momentum distribution of the hole, $n^h(\mathbf{k})$%
, and the quasiparticle weight $Z_{\mathbf{k}}$, as calculated by VMC in the
ground state ansatz of Eq. (\protect\ref{gs}): (a) and (b) show excellent
agreement with the DMRG on a $8\times 8$ lattice; (c) and (d): a larger size
at $16\times16$. The finite-size scaling results obtained by VMC are
presented in Fig. \protect\ref{Fig2}.}
\label{t-Jfour}
\end{figure*}

Such a unique ansatz in Eq.~(\ref{gs}) can be compared to the following
Bloch-like one-hole state 
\begin{equation}  \label{Bloch}
|\Psi _{\text{Bloch}}\rangle_{\text{1h}}\propto \sum_{i} e^{i{\mathbf{k}}%
\cdot {\mathbf{r}}_i} {c}_{i\downarrow }|\text{RVB}\rangle ~.
\end{equation}
Besides the momentum $\mathbf{k}$, the state of Eq.~(\ref{Bloch}) carries
a total spin $S=1/2$, $S^z=1/2$, and a charge $+e$, which is created by the
bare hole operator ${c}^{}_{i\downarrow }$ on a spin-singlet and
translationally invariant spin background. So the new ground state of Eq.~(%
\ref{gs}) means that the ``quasiparticle'' creation operator is changed to 
\begin{equation}
{c}_{i\sigma}\rightarrow \tilde{c}_{i\sigma}~,
\end{equation}
or equivalently the single-hole wave function is changed from a Bloch wave to
a many-body version by 
\begin{equation}
e^{i{\mathbf{k}}\cdot {\mathbf{r}}_i} \rightarrow \varphi _{\text{h}}\left(
i\right)e^{-i\hat{\Omega}_i}~.
\end{equation}
Indeed, the new quasiparticle, created by $\tilde{c}$ in Eq.~(\ref{gs}),
can propagate more coherently as compared to the bare $c$ in the
antiferromagnetic spin background (cf. Fig.~\ref{corr}). It is noted that
the ansatz state in Eq.~(\ref{gs}) is defined in a finite-size system with
open boundary condition (OBC), in which the $C_4$ rotational symmetry is
retained. Besides the total spin $S=1/2$, $S^z=1/2$, and the hole number $N_%
\text{h}=1$, it shows a nontrivial angular momentum $L_z=\pm 1$ in agreement
with the ED and DMRG \cite{Zheng2018}, indicating that there is a novel
double ground state degeneracy for a given $S^z$. Such a nontrivial angular
momentum $L_z=\pm 1$ is shown to be associated with the neutral spin current
pattern and charge current pattern, respectively, in Figs.~\ref{fig:Js} and %
\ref{fig:Jh}. These currents are qualitatively consistent with the ED and
DMRG results \cite{Zheng2018}, which can be directly connected to the phase
shift factor $e^{-i\hat{\Omega}_i}$ in the ground state of Eq.~(\ref{gs}) as
schematically illustrated in Fig.~\ref{Fig1}(a).

As shown in Fig.~\ref{Fig1}(b), the ground state of Eq.~(\ref{gs}) is
further composed of four Bloch-wave states [Eq.~(\ref{Bloch})] characterized
by the quasiparticle spectral weight $Z_{\mathbf{k}}$, which is peaked at
four momenta 
\begin{equation}  \label{K0}
\mathbf{K}^0=(\pm \frac{\pi}{2}, \pm \frac{\pi}{2}) ~.
\end{equation}
$Z_{\mathbf{k}}$ measures the quasiparticle spectral weight, which shows
that the ground state of Eq.~(\ref{gs}) automatically includes four Bloch
wave components at momenta $\mathbf{K}^0$. This is consistent with the ED
calculation where the four-fold degeneracy with the total momenta $\mathbf{K}%
^0$ has been identified on a torus (PBC) at a large ratio of $t/J$ in the $t$%
-$J$ model\cite{Zheng2018}. In particular, the value of $Z_{\mathbf{k}}$
calculated by VMC agrees very well with that computed by DMRG for a lattice
size $N=8\times 8$ under OBC as shown in Fig.~\ref{t-Jfour}(b) at $t/J=3$.

\begin{figure}[tb]
\begin{center}
\includegraphics[width=0.5\textwidth]{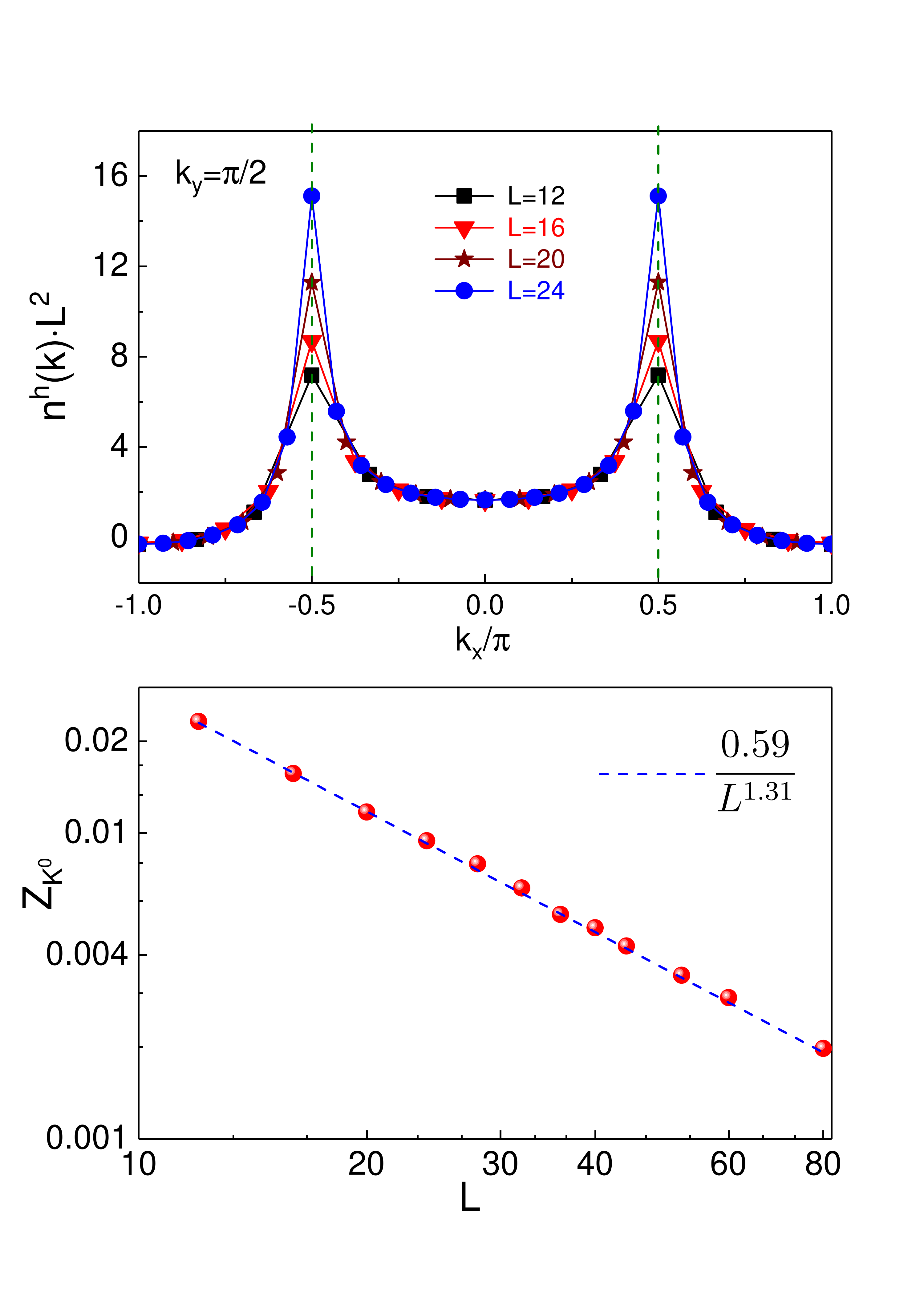}
\end{center}
\par
\renewcommand{\figurename}{Fig.} 
\caption{(Color online.) The scaling analysis of the momentum distribution $%
n^h\left( \mathbf{k}\right) $ and the quasiparticle weight $Z_{\mathbf{K}^0}$%
. (a) The broad background part of $n^h\left( \mathbf{k}\right) $ scales
with the inverse of $L^2$ in the square sample $N=L\times L$; (b) The
quasiparticle weight $Z_{\mathbf{K}^0}$ at the peak momentum $\mathbf{K}%
^0=(\pm\frac{\protect\pi}{2},\pm\frac{\protect\pi}{2})$ vanishes in a
power-law fashion by $L^{-1.31}$. }
\label{Fig2}
\end{figure}

Figure \ref{t-Jfour}(d) further shows the VMC result of $Z_{\mathbf{k}}$ at $%
N=16\times 16$. In fact, at larger sample sizes, $Z_{\mathbf{k}}$ is shown
to vanish in a power-law fashion (cf. Fig. \ref{Fig2}(b)) as follows: 
\begin{equation}  \label{Zk}
Z_{\mathbf{K}^0}\simeq 0.59 \frac{1}{L^{1.31}}
\end{equation}
at $t/J=3$ with $L$ denoting the sample length. In other words, the ground
state ansatz (\ref{gs}) is a prototypic non-Fermi liquid state with
vanishing $Z_{\mathbf{k}}$ at $N\equiv L^2 \rightarrow \infty$. Such a
``twisted'' quasiparticle is non-Landau-like, in contrast to the
conventional Landau quasiparticle implied in the Bloch-wave state (\ref%
{Bloch}).

Furthermore, the momentum structure of the doped hole can be also measured
by the hole momentum distribution function defined by 
\begin{equation}  \label{nh}
n^h(\mathbf{k})\equiv 1-n^e(\mathbf{k})=1-\left \langle \sum_{\sigma}{c}%
^{\dagger}_{\mathbf{k}\sigma}{c}_{\mathbf{k}\sigma}^{}\right\rangle~,
\end{equation}
where $n^h(\mathbf{k})=0$ at half-filling and $\sum_{\mathbf{k}}n^h(\mathbf{k%
})=1$ for the one hole case. The good agreement between the VMC calculation
based on Eq. (\ref{gs}) and the DMRG result at $N=8\times 8$ can be found in
Fig.~\ref{t-Jfour}(a). Figure \ref{t-Jfour}(c) further shows the VMC result
at $N=16\times 16$. A finite-size scaling is presented in Fig. \ref{Fig2}%
(a), which indicates that besides the peaks at $\mathbf{K}^0$ proportional
to $Z_{\mathbf{K}^0}$, $n^h(\mathbf{k})$ exhibits a broad continuum, which
satisfies a scaling $\propto 1/N$ and thus its total weight contributes to a
finite and predominant part to the sum rule of $\sum_{\mathbf{k}}n^h(\mathbf{%
k})$. By contrast, the quasiparticle component vanishes as given in Eq. (\ref%
{Zk}), such that the bare hole truly becomes incoherent in the thermodynamic
limit  demonstrated  by extrapolation of finite size results.

Therefore, the one-hole ground state (\ref{gs}) is composed of two
components, in which the doped hole either behaves like a Bloch wave at the
four Fermi points of $\mathbf{K}^0$ with the spectral weight $Z_{\mathbf{K}%
^0}$ or becomes incoherent with a broad momentum distribution. In the later
component, a partial momentum is carried away by the \emph{neutral} spin
currents as presented in Fig.~\ref{fig:Js}. The Landau's one-to-one
correspondence hypothesis no longer holds true here in the presence of the
second component, which violates the adiabaticity by allowing a continuous
momentum transfer between the hole and the surrounding spin background. In
particular, $\varphi _{\text{h}}$ determined by VMC is no longer
Bloch-wave-like (cf. Fig.~\ref{Fig:BlochEnergy}(b)), which means the translational symmetry is explicitly broken.

Finally, we find that the Bloch-like ground state in Eq. (\ref{Bloch}) will
replace Eq. (\ref{gs}) to become the ground state variationally for the $%
\sigma\cdot t\text{-}J$ model. All the novel properties including the double
ground state degeneracy, nontrivial $L^z$ with finite spin/charge currents
surrounding the hole/spin, vanishing $Z_{\mathbf{k}}$, and the two-component
feature, etc., disappear in such a ground state. In other words, we find that the
Landau's quasiparticle description for the doped hole is recovered in the
model in which the phase string effect is turned off in the hopping term,
while the superexchange term remains unchanged. Thus the present variational
approach clearly establishes that the phase string effect hidden in the 2D $t
$-$J$ model is well encoded by the phase string operator in Eq.~(\ref{Omega}%
), which gives rise to the mutual spin/charge currents as the composite
structure associated with the doped hole that is ``twisted'' according to
Eq.~(\ref{tildec}) in an antiferromagnetic background $|\text{RVB}\rangle $.

\section{Ground state ansatz and variational Monte Carlo calculation}

\label{Sec::Gs ansatz}

\subsection{Variational ground state for the one-hole doped Mott insulator}

The single-hole-doped ground state may be generally constructed by removing
an electron with, say, $\downarrow$ spin, from the half-filling ground state 
$|\text{RVB} \rangle$ as follows 
\begin{equation}  \label{eq:psiG}
|\Psi _{\text{G}}\rangle _{\text{1h}}=\sum_{i}\left(\varphi _{\text{h}}(i) 
\hat{\Pi}_i\right)c_{i\downarrow }|\text{RVB} \rangle,
\end{equation}
where the summation is over the lattice site $i$ weighted by a hole
wave function $\varphi _{\text{h}}(i)$ and a many-body operator $\hat{\Pi}_i$%
, which denotes the generic distortion (i.e., spin polaron) of the spin
background in response to the injection of a hole into the half-filling
ground state.

The ground state of the Heisenberg type Hamiltonian $H_J$ at half-filling is
denoted by $|\text{RVB} \rangle$ above. So far the best variational
wave function is the so-called \emph{bosonic} resonant valence bond (RVB)
state proposed \cite{Liang1988some} by Liang, Doucot, and Anderson, which is
a spin singlet with  translational invariance, and has a very accurate
variational energy [$E_{\mathrm{G}}^0/(2N)+1/4J=-0.3344J$] as compared to the precise
numerical results. Our VMC approach will be based on such an $|\text{RVB}
\rangle$ as the starting point (Appendix A).

If one neglects the ``spin polaron'' effect of $\hat{\Pi}_i$ in Eq.~(\ref%
{eq:psiG}) by taking $\hat{\Pi}_i=1$, a Bloch-like state $%
|\Psi_{\mathrm{Bloch}}\rangle_{\mathrm{1h}}$  will be reproduced with 
\begin{equation}  \label{bwave}
\varphi _{\text{h}} (i) \propto e^{i{\mathbf{k}}\cdot {\mathbf{r}}_i}~,
\end{equation}
as given in Eq.~(\ref{Bloch}), which describes a Landau's quasiparticle
with the quantum numbers of total spin $S=1/2$, $S^z=1/2$, charge $+e$, and
a momentum $\mathbf{k}$ corresponding to the translational operation by a
distance ${\mathbf{l}}$, 
\begin{equation}  \label{t}
\hat{T}_{\mathbf{l}}|\Psi_{\mathrm{Bloch}} \rangle _{\text{1h}}=e^{-i{\mathbf{%
k}}\cdot {\mathbf{l}}}|\Psi_{\mathrm{Bloch}} \rangle _{\text{1h}} ~,
\end{equation}
by noting $\hat{T}_{\mathbf{l}} |\text{RVB} \rangle=|\text{RVB} \rangle$, $%
\hat{T}_{\mathbf{l} } c_{i\downarrow }\hat{T}_{\mathbf{l} }^{\dagger}= c_{i+%
\mathbf{l}\downarrow } $.

Such a Bloch wave picture, marked by Eq.~(\ref{bwave}), would remain
unchanged at $\hat{\Pi}_i\neq 1$ if one requires the translational symmetry (%
\ref{t}) under a many-body operator $\hat{T}_{\mathbf{l}}$ involving the
hole and \emph{whole} spins. (Note that $\hat{\Pi}_i$ generally satisfies
the translational symmetry by $\hat{T}_{\mathbf{l} } \hat{\Pi}%
_ic_{i\downarrow }\hat{T}_{\mathbf{l} }^{\dagger}= \hat{\Pi}_{i+\mathbf{l}%
}c_{i+\mathbf{l}\downarrow } $.) However, in the present variational scheme, 
$\varphi _{\text{h}} (i)$ obtained by VMC with $\hat{\Pi}_i\neq 1$ may not
necessarily recover the Bloch-wave solution of Eq.~(\ref{bwave}). In other
words, a spontaneous translational symmetry breaking may occur in the
single-hole ground state (\ref{eq:psiG}), as to be shown below.

First, we note that the Bloch-wave state indeed holds true generally if $%
\hat{\Pi}_i$ represents a \emph{local} spin distortion rigidly bound to the
hole, solely specified by the hole or a ``centre-of-mass'' coordinate via
the hole wave function $\varphi _{\text{h}}(i)$. Indeed, the ``longitudinal
spin polaron''\cite{Schmitt-Rink1988Spectral, Kane1989,Martinez1991spin,
Liu1991Spectral} or ``spin bag'' effect\cite{Schrieffer1988Spin, Weng1988d}
can generally improve the ground state energy of the Bloch state (\ref{Bloch}%
) without changing its nature as a Landau's quasiparticle, except for a
renormalization of the effective mass or even the location of $\mathbf{k}$
in the ground state.

\begin{figure}[h]
\begin{center}
\includegraphics[width=0.5\textwidth]{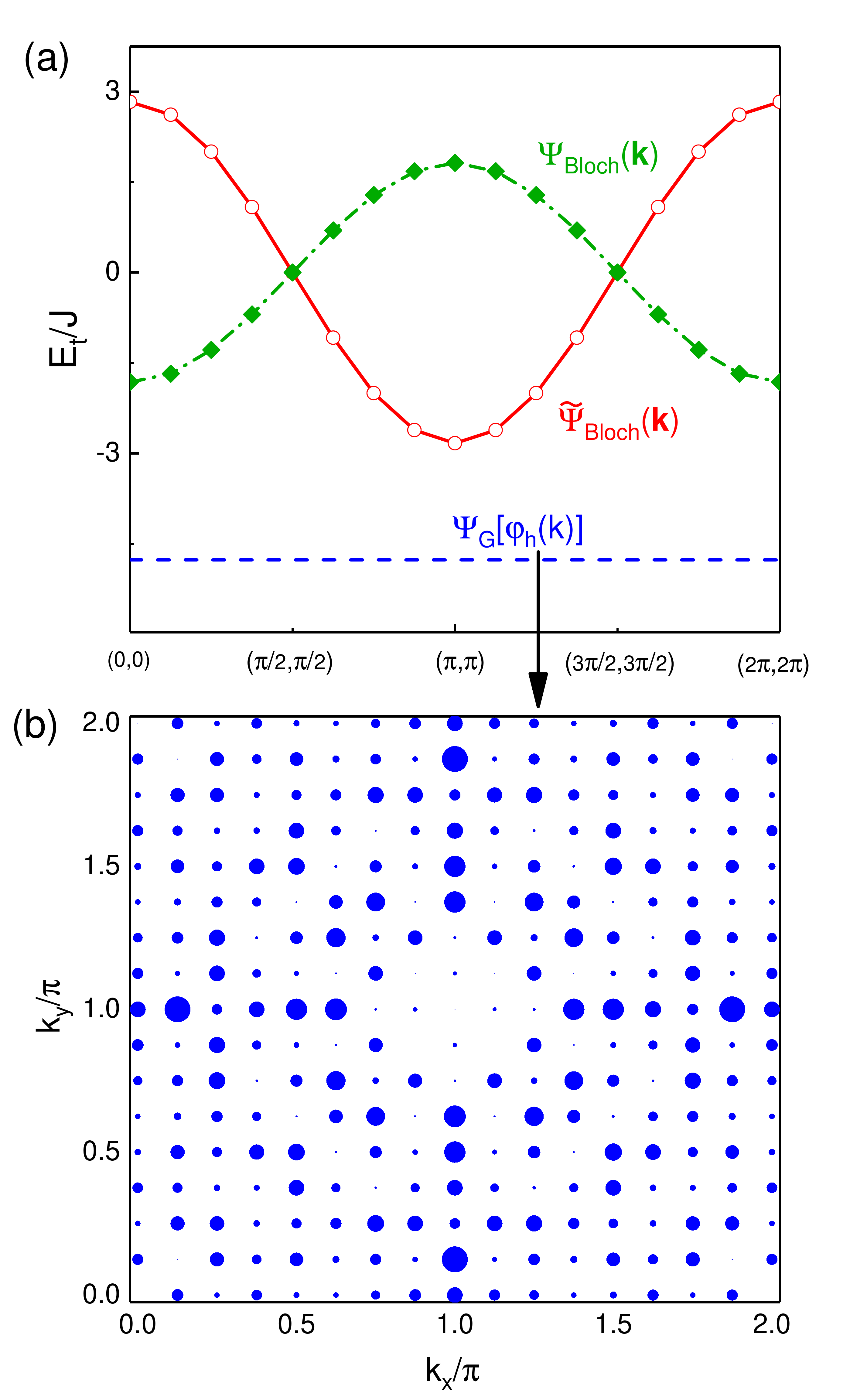}
\end{center}
\par
\renewcommand{\figurename}{Fig.}
\caption{(Color Online.) (a) The kinetic energies of the Bloch-like states $|\Psi_{\mathrm{Bloch}}(\mathbf{k})\rangle_{\mathrm{1h}}$ [Eq.~(\ref{Bloch})] and $|\widetilde{\Psi}_{\mathrm{Bloch}}(\mathbf{k})\rangle_{\mathrm{1h}}$ [Eq.~(\ref{Bloch-1})] as a function of momentum vs. that of the variational ground state $|\Psi_{\mathrm{G}}\rangle_{\mathrm{1h}}$ in Eq.~(\ref{gs}) (dashed line); Note that the hole wave function $\varphi_{\mathrm{h}}(i)$ determined variationally in Eq.~(\ref{gs}) is not translational invariant as shown in (b); (b) The distribution of the absolute value of $\varphi_{\mathrm{h}}(\mathbf{k})$ as the Fourier transformation of $\varphi_{\mathrm{h}}(i)$ in the momentum space. 
 }
\label{Fig:BlochEnergy}
\end{figure}

However, in this work, we shall focus on a new type of ``transverse spin
polaron'' effect in $\hat{\Pi}_i$ given by \cite{Weng2011superconducting,
Weng2011mott} 
\begin{equation}  \label{vwf}
\hat{\Pi}_i=e^{-i\hat{\Omega}_i}~,
\end{equation}%
where the many-body phase shift operator $\hat{\Omega}_i$ is defined in Eq.~(%
\ref{Omega}), which will introduce \emph{transverse spin currents} around
the hole (see below). The precise form of Eq.~(\ref{Omega}) is originated
from the intrinsic phase string sign structures\cite{Weng1996Phase,
Weng1997Phase, Wu2008Sign} of the $t$-$J$ model, representing a long-range
mutual entanglement between the doped hole and the spin background.
Physically, the phase shift in Eq.~(\ref{vwf}) describes a nonlocal response
of the \emph{whole} spin degree of freedom to the injection of a hole, which
is non-perturbative in nature. Its explicit expressions in one-dimension\cite%
{Wang2014Sign, Zhu2016Exact} and two-leg ladder\cite{Wang2015variational,
Chen2018two} systems have been previously studied in earlier works.

The VMC simulation (cf. Appendix B) shows that the ground state energy of the ansatz state (%
\ref{gs}) will be lowered as compared to that of the Bloch state (\ref{Bloch}%
) by $\Delta E_{\mathrm{G}}=-1.50J$ at $N=16\times16$ (in which the kinetic energy
difference is $\Delta E_t=-2.71J$ and the superexchange energy difference is 
$\Delta E_J=1.21J$).  Here, the absolute value of $\varphi_{\text{h}}(\mathbf{k})$, the Fourier transformation of $\varphi_{\text{h}}(i)$ determined by optimizing the ground state energy, is shown in Fig.~\ref{Fig:BlochEnergy}(b).  Clearly it is not solely peaked at four $\mathbf{K}^0$ in Eq.~(\ref{K0}) as would be 
expected for a linear superposition of four translational invariant states
with total momenta $\mathbf{K}^0$, which are to be obtained later by calculating the quasiparticle spectral weight. 

In particular, if one assumes a translationally invariant form with $\varphi_{\text{h}}(i)$ taking the Bloch-wave form  in Eq.~(\ref{bwave}) 
\begin{equation}  \label{Bloch-1}
|\widetilde{\Psi }_{\text{Bloch}}(\mathbf{k})\rangle_{\text{1h}}\propto \sum_{i} e^{i{\mathbf{k}}%
\cdot {\mathbf{r}}_i} \tilde{c}_{i\downarrow }|\text{RVB}\rangle ~,
\end{equation}
the resulting kinetic energy is  \emph{higher} by $\Delta E_t= 1.94J$ at $N=16\times 16$ (the superexchange energy is the same). Therefore, in contrast to Eq.~(\ref{Bloch-1}), the wave function $\varphi_{\text{h}}(i)$ as determined variationally in Eq. (\ref{gs}) indeed automatically breaks the translational symmetry in the true variational ground state.

It is noted that the VMC calculation for an $8\times 8$ square lattice with OBC ($t=3J$) gives rise to a kinetic energy $E_t=-5.08J$ and the superexchange energy $E_J=-37.21J$ (at half-filling $E_J=-39.4884J$), while the DMRG gives the values of $E_t=-8.67J$ and  $E_J=-37.28J$. Here we emphasize that one may further improve the single hole’s kinetic energy by incorporating the SCBA-like correction into the ansatz state (\ref{gs})  without changing the nature of the composite/fractionalization structure of the one-hole ground state. But the absolute kinetic energy is not our main concern here. Instead, we shall focus more on the structure and ground state properties of the ansatz state (\ref{gs}) in comparison with the DMRG simulation.

On the other hand, as we shall see later, with the phase string being switched off
in the so-called $\sigma\cdot$ $t$-$J$ model by restoring the trivial sign
structure (cf. Sec.~\ref{tj-stj}), the one-hole wave function will simply
reduce to the Bloch wave form in Eq.~(\ref{Bloch}) with a lower ground state
energy than that of Eq. (\ref{gs}). Thus, the phase shift operator of Eq.~(\ref{Omega}) is really
originated from the phase string, which must be ``turned off'' in Eq. (\ref%
{eq:psiG}) in the absence of such an effect. Consequently $\varphi_{\text{h}%
}(i)$ restores the Bloch wave form in Eq.~(\ref{bwave}).

Although one may further improve the one-hole ground state energy by
incorporating the ``longitudinal spin polaron'' effect\cite%
{Shraiman1988mobile, Shraiman1990mobile,Schmitt-Rink1988Spectral, Kane1989,
Martinez1991spin,Liu1991Spectral} mentioned above for both $t$-$J$ and $%
\sigma\cdot$ $t$-$J$ models, in the present work, our main focuses will be
on the qualitatively different properties exhibited between the ground state
of Eq.~(\ref{gs}) and the Bloch-wave state in Eq.~(\ref{Bloch}), and the
conventional ``longitudinal spin polaron'' effect will be omitted for the
sake of simplicity.

\subsection{Hidden spin/charge currents}

It can be explicitly seen that the ground state ansatz $|\Psi _{\text{G}%
}\rangle _{\text{1h}}$ in Eq.~(\ref{gs}) has the hole number $N_{\text{h}}=1$
and total spin $S=S^z=1/2$, which are conserved in the $t$-$J$ Hamiltonian.
In the Heisenberg picture, the corresponding continuity equations associated
with these quantities are given as follows \cite{Zheng2018} 
\begin{align}
\frac{\mathrm{d }n_i^h}{\mathrm{d}\tau} &= \sum_{j=\mathrm{NN}(i)} J_{ij}^h~,
\\
\frac{\mathrm{d }S_i^z}{\mathrm{d}\tau} &= \sum_{j=\mathrm{NN}(i)} (J_{ij}^s
+ J_{ij}^b)~,
\end{align}
where $\tau$ denotes the time, $J^h$ the hole (charge) current, $J^b$ the
backflow spin current associated with the hole hopping, and $J^s$ the
neutral spin current in the Heisenberg background, which are respectively
defined by 
\begin{align}  \label{eq:Js}
J_{ij}^{{h}} &=-it\sum_{\sigma }\left( c_{i\sigma }^{\dag }c_{j\sigma
}^{{}}-c_{j\sigma }^{\dag }c_{i\sigma }^{{}}\right)~, \\
J_{ij}^{{b}} &=i\frac{t}{2}\sum_{\sigma }\sigma \left( c_{i\sigma }^{\dag
}c_{j\sigma }^{{}}-c_{j\sigma }^{\dag }c_{i\sigma }^{{}}\right)~, \\
J_{ij}^{{s}} &=-\frac{J}{2}i\left(
S_{i}^{+}S_{j}^{-}-S_{i}^{-}S_{j}^{+}\right)~.
\end{align}

\begin{figure}[th]
\begin{center}
\includegraphics[width=0.35\textwidth]{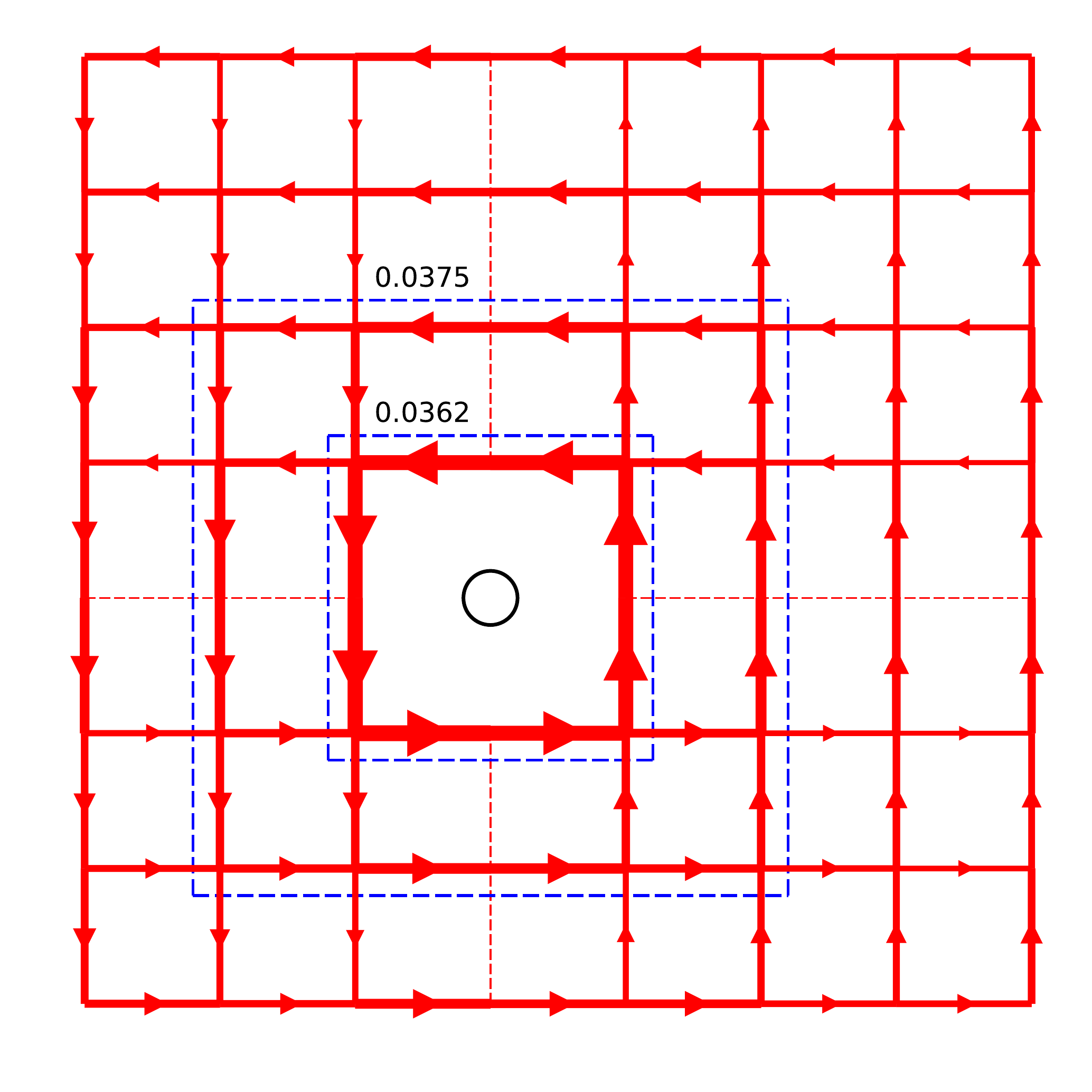}
\end{center}
\par
\renewcommand{\figurename}{Fig.} 
\caption{(Color Online.) The neutral spin current ($J^s_{ij}$) pattern
surrounding the hole in one of the degenerate ground states with $L_z=1$ in
a lattice of $N=8\times 8$ with OBC. (The red vertical and horizontal dashed
lines mark the bonds with vanishing spin currents.) The dashed blue closed
loops and numbers indicate the Wilson loops which count the Berry phase
defined in Eq.~(\protect\ref{W}). }
\label{fig:Js}
\end{figure}

\begin{figure}[th]
\centering
\includegraphics[width=0.35\textwidth]{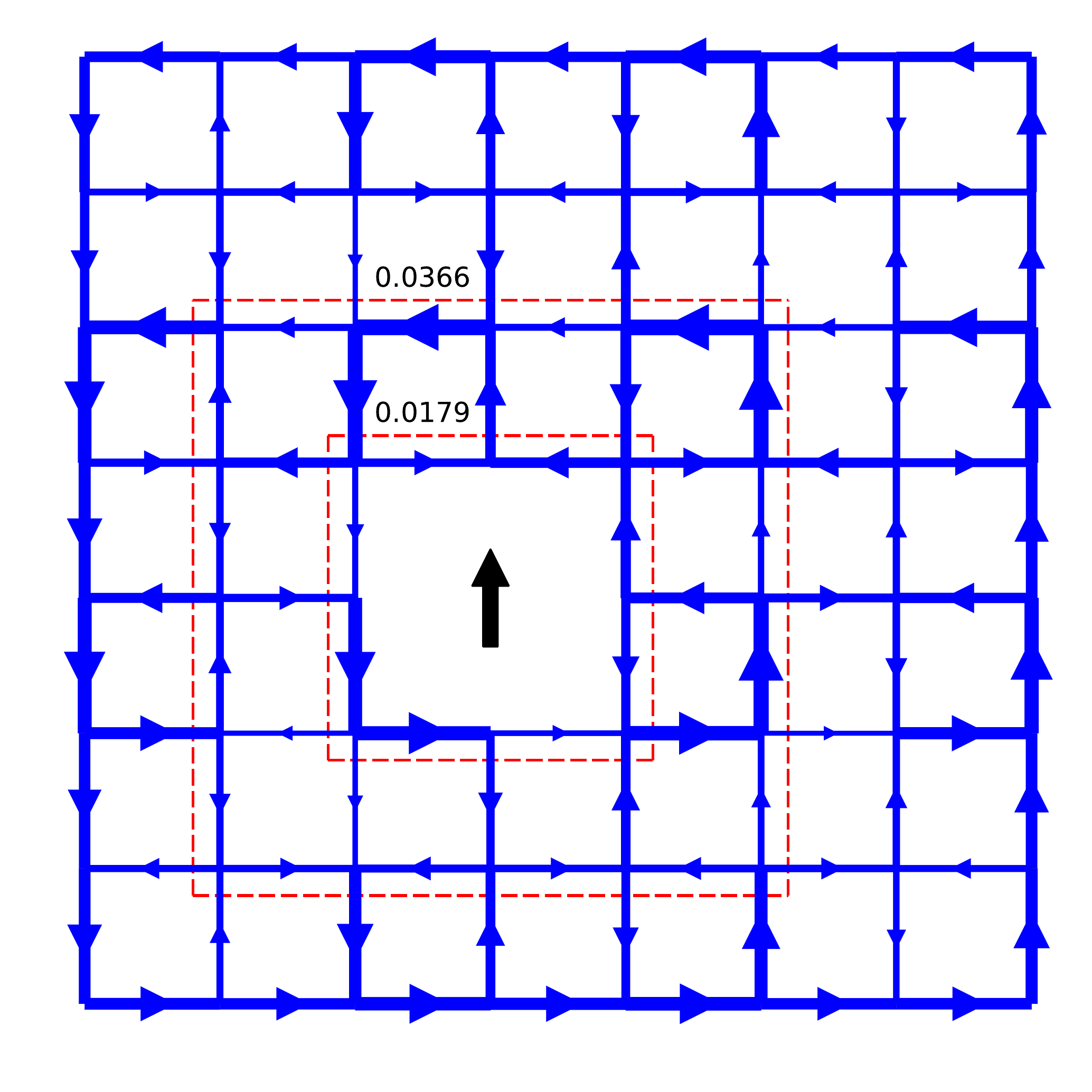} %
\renewcommand{\figurename}{Fig.} 
\caption{(Color Online.) The charge (hole) current ($J^h_{ij}$) pattern
surrounding an $\uparrow$ spin projected onto a given lattice site in a degenerate ground state with $L_z=1$ in a
lattice of $N=8\times 8$ with OBC. The dashed red closed loops and numbers
indicate the Wilson loops counting the Berry phase defined in Eq.~(\protect
\ref{T}). }
\label{fig:Jh}
\end{figure}

By using VMC, one can variationally determine the single hole wave function $%
\varphi _{\text{h}}\left(i\right) $ of the ansatz state (\ref{gs}). The
corresponding instant patterns of spin currents around the hole and the hole
currents around an $\uparrow$ spin are obtained by computing the following
correlation functions: 
$\langle \mathcal{P}_l^h J_{ij}^s \rangle $ and $\langle
\mathcal{P}^s_m J_{ij}^h \rangle$  where $\mathcal{P}_l^h\equiv n_{i_0}^h$ and $\mathcal{P}^s_l \equiv c_{l\uparrow}^{\dagger} c_{l\uparrow}^{}$ project
the hole and an $\uparrow$ spin at site $l$, as shown in Figs.~\ref{fig:Js}
and \ref{fig:Jh}, respectively. In Fig.~\ref{fig:Js}, circulating neutral
spin currents surrounding the doped hole are clearly shown, and mutually
hole currents circulating a fixed ($\uparrow$) spin are presented in Fig.~%
\ref{fig:Jh}. Their chirality depends on the sign of $\hat\Omega_i$ in Eq.~(%
\ref{Omega}), which is concomitant with a double degeneracy of the ground
state specified by an angular momentum $L_z=\pm 1$ to be discussed later.


These novel spin and charge currents can be directly traced back to the
phase shift operator in Eq. (\ref{vwf}), which can be associated with a
nontrivial Berry phase \cite{Weng2001spin}. Generally, the Berry phase of
Eq. (\ref{gs}) can be identified by its phase change accumulated under an
adiabatic change of the wave function along a closed loop in some parameter
space. Here it is specified by the space-time path of the hole and spin
configurations. 

If we examine a loop in the parameter space describing the full
braiding between such a spin and the doped hole, one would find two Berry
phases, one corresponding to the winding of the spin at $m$ around the hole
at $i_0$ via a closed loop $c$ encircling but not crossing $i_0$ 
\begin{equation}   \label{W}
\gamma^{h-s} _{i_{0}}\left[ c\right]\propto W\left[ c\right] \equiv
\left\langle \oint_{c} \mathcal{P}_{i_0}^{h} J_{ij}^{\text{s}} \right \rangle,
\end{equation}
and the other the winding of the hole around the spin, given by 
\begin{equation}  \label{T}
\gamma^{s-h} _{m}\left[ c\right]\propto T\left[ c\right] \equiv \left\langle
\oint_{c} \mathcal{P}_m^{s} J_{ij}^{\text{h}} \right \rangle.
\end{equation}%
Both nontrivial $\gamma^{h-s} _{i_{0}}\left[ c\right]$ and $\gamma^{s-h} _{m}%
\left[ c\right]$ are thus directly connected with the spin and charge
current loops shown in Figs.~\ref{fig:Js} and \ref{fig:Jh}.

\subsection{Nontrivial quantum number: Angular momentum $L_z=\pm 1$}

\label{Appen:Lz}

Let us start with a system of the square lattice of a finite size of $%
2M\times 2M$, which possesses a $C_4$ rotational symmetry under the OBC. A
straightforward manipulation based on the wave function in Eq.~(\ref{gs}) can
demonstrate (see below) that under a spatial rotation of $\pi/2$, the ground
state will be transformed by 
\begin{equation}  \label{R}
\hat{R}(\pi/2) |\Psi _{\text{G}}\rangle _{\text{1h}} = \pm i |\Psi _{\text{G}%
}\rangle _{\text{1h}}~,
\end{equation}
where $\hat{R}(\theta)$ is the spatial rotational operator of angle $\theta$
with eigenvalue $e^{iL_z\theta}$. So the ground state has a nonzero angular
momentum $L_z=\pm 1$ and a precise two-fold degeneracy under a given $S^z$,
which are in agreement with the numerical result \cite{Zheng2018} for
finite-size systems with $t/J=3$.

The proof of Eq.~(\ref{R}) is given as followed. Let $\hat{R}\equiv \hat{R}(\pi/2)$ be the
operator that rotates the system anticlockwisely by 90 degrees. When acting $%
\hat{R}$ on the variational wave function (\ref{gs}), one has 
\begin{eqnarray}  \label{eq:symm:wf}
\hat{R}|\Psi_{\text{G} }\rangle_{\text{1h}} &=&\hat{R}\sum_{i}\varphi _{\mathrm{h}%
}\left( i\right) e^{-i\hat{\Omega}_i}c_{i\downarrow }|\mathrm{RVB}\rangle  \notag
\\
&=&\sum_{i}\varphi _{\mathrm{h}}\left( i\right) \left( \hat{R}e^{-i%
\hat{\Omega}_i}c_{i\downarrow }\hat{R}^{-1}\right) \hat{R}|\mathrm{RVB}\rangle~.
\end{eqnarray}%
The spin RVB state by construction is rotationally invariant 
\begin{equation}
\hat{R}|\mathrm{RVB}\rangle =|\mathrm{RVB}\rangle~.
\end{equation}%
The symmetry transformation of the combination of the phase string operator
and the fermion annihilation operator is 
\begin{eqnarray}
\hat{R}e^{-i\hat{\Omega}_i}c^{}_{i\downarrow }\hat{R}^{-1} &=&\hat{R}\exp\left( \pm i\sum_{l\not=i}\theta _{i}\left( l\right) n^{}_{l\downarrow
}\right) c^{}_{i\downarrow }\hat{R}^{-1}  \notag \\
&=&\exp \left( \pm i\sum_{l\not=i}\theta _{i}\left( 
 l\right) n^{}_{\hat{R}l\downarrow }\right) c^{}_{\hat{R}i\downarrow }~, 
\notag
\end{eqnarray}%
which can be further simplified by using $\theta^{}_{\hat{R}i}\left( \hat{R}%
l\right) =\theta _{i}\left( l\right)+\pi /2$ and $\sum_{l(\not=i)}n^{}_{l\downarrow }=\sum_{l(\not= i)}\left( \frac{n^{}_{l\downarrow
}+n^{}_{l\uparrow }}{2}-\frac{n^{}_{l\uparrow }-n^{}_{l\downarrow }}{2}\right) =\frac{%
N-1}{2}-S^z$. Therefore, the symmetry transformation of the wave function
Eq.~(\ref{eq:symm:wf}) becomes 
\begin{eqnarray}
& &\hat{R}|\Psi_{\text{G} }\rangle_{\text{1h}}  \notag \\
&=& \sum_{i}\varphi^{} _{\mathrm{h}}\left( i\right) \exp \left[ \pm i\frac{%
\pi }{2}\left( \frac{N-1}{2}-S^{z}\right) \right] e^{-i\hat{\Omega}_i}
c_{i\downarrow }|\text{RVB}\rangle  \notag \\
& =& \exp \left[ \pm i\frac{\pi }{2}\left( \frac{N-1}{2}-S_{z}\right) \right]
|\Psi_{\text{G}}\rangle_{\text{1h}}~.  \label{Lzphase}
\end{eqnarray}
For a bipartite lattice with size $N=2M\times 2M$ and $S^z=1/2$, one finds
that Eq.~(\ref{R}) holds true. On the other hand, for a lattice with odd
number of total sites, for example, $N=5\times 5$, the phase factor in Eq.~(%
\ref{Lzphase}) is $\pm 1$. All these are consistent with ED and DMRG
simulations \cite{Zheng2018}.

\subsection{Equal-time single-hole propagation}

\begin{figure}[tb]
\includegraphics[width=0.5\textwidth]{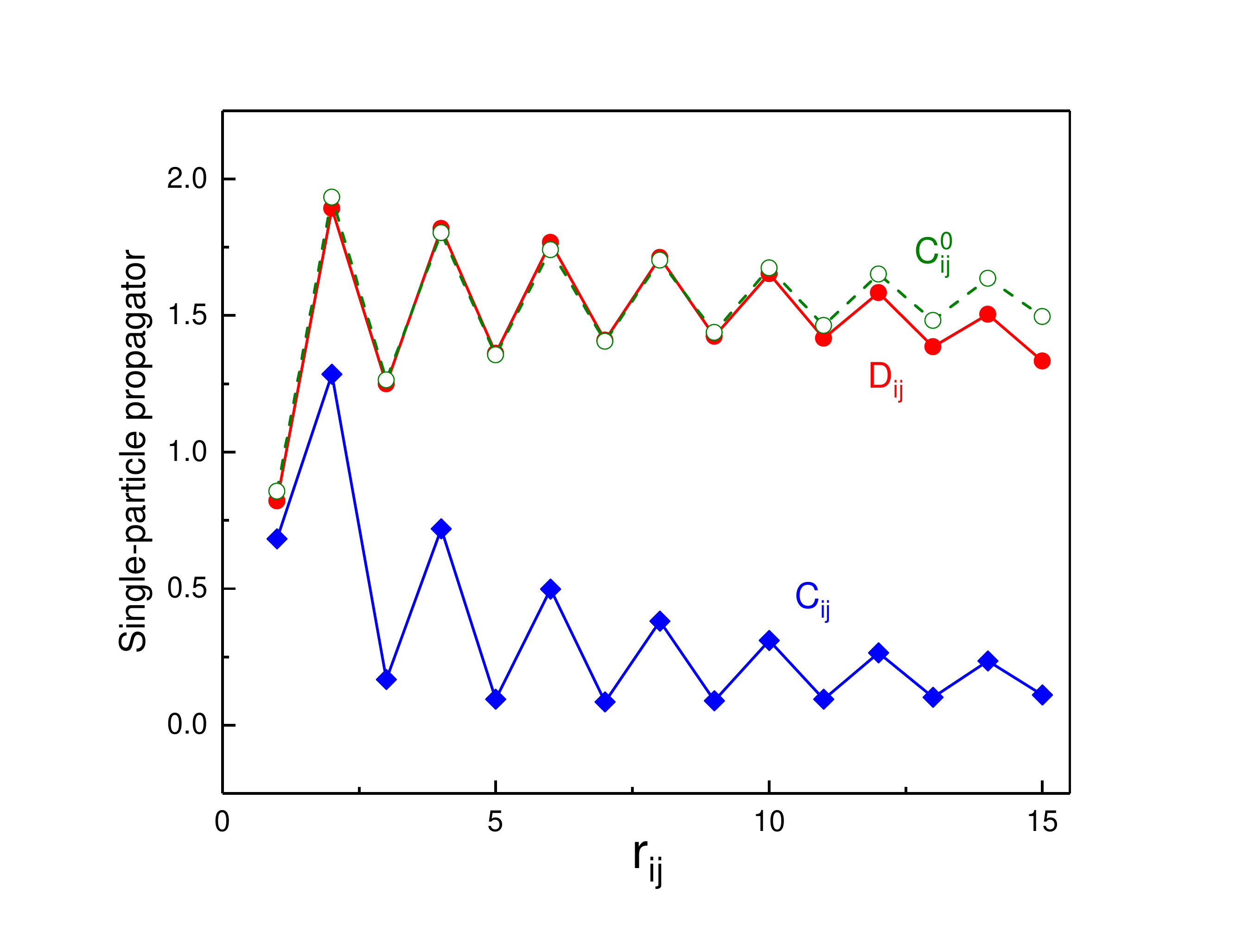} 
\par
\renewcommand{\figurename}{Fig.}
\caption{(Color online.) The propagation amplitude of a bare hole, $\mathcal{C}_{ij}$,
is much reduced as compared to that of the ``twisted'' hole, $\mathcal{D}_{ij}$, in
the ground state ansatz (\protect\ref{gs}). The latter is comparable to the
hole propagation $\mathcal{C}^{0}_{ij}$ for a Bloch wave state defined in Eq.~(\protect
\ref{Bloch}) (with momentum $\mathbf{K}^0$). Here the spatial distance $%
r_{ij}=|x_i-x_j|+|y_i-y_j|$, and ${\mathcal{C}}_{ij}$ and ${\mathcal{D}}_{ij}$ are calculated by
averaging over all the lattice sites with each given distance $r_{ij}$ based
on the definitions given in Eq.~(\protect\ref{C}) and Eq.~(\protect\ref{D}), respectively.
The lattice size is $20\times20$. }
\label{corr}
\end{figure}

In the single-hole ground state, one may examine the single-hole propagator
defined by 
\begin{equation}  \label{C}
\mathcal{C}_{ij}=\left\langle \Psi _{\text{G}}\right| c_{j\downarrow
}^{}c_{i\downarrow }^{\dagger }\left|\Psi _{\text{G}}\right\rangle _{\text{1h%
}}~.
\end{equation}
The result calculated by VMC is presented in Fig.~\ref{corr}. We can see
that the propagator $\mathcal{C}_{ij}$ is substantially suppressed and
decays much faster than in a conventional Bloch wave state of Eq.~(\ref
{Bloch}) (with the momentum at $\mathbf{K}^0$ for the convenience of
comparison) as shown by $\mathcal{C}^0_{ij}$ in Fig.~\ref{corr}, which is defined by
\begin{equation}
\mathcal{C}_{ij}^0=\left\langle\Psi_{\mathrm{Bloch}}\right
|c^{}_{j\downarrow}c^{\dagger}_{i\downarrow}\left|\Psi_{\mathrm{Bloch}%
}\right\rangle_{\mathrm{1h}}~.
\end{equation}
It implies that the ground-state ansatz in Eq.~(\ref{gs}) does not favor a coherent
propagator of a bare hole on the quantum spin background.

We have seen that in the ground state ansatz (\ref{gs}), the doped hole is
``twisted'' into a composite hole described by $\tilde{c}_{i\downarrow }$ in
Eq. (\ref{tildec}). Thus, a new hole object characterized by $\tilde{c}%
_{i\downarrow }$ is expected to propagate more coherently in the following
propagator: 
\begin{eqnarray}  \label{D}
\mathcal{D}_{ij} &=&\left\langle \Psi _{\text{G}}\right| \tilde{c}%
_{j\downarrow }\tilde{c}_{i\downarrow }^{\dagger }\left|\Psi _{\text{G}%
}\right\rangle_{\text{1h}}  \notag \\
&=&\varphi^{*}_{\mathrm{h}}(j)\varphi^{}_{\mathrm{h}}(i)\left\langle
\!(\frac{1}{2}-S^z_j)(\frac{1}{2}%
-S^z_i)\right\rangle_{\mathrm{RVB}}~,
\end{eqnarray}%
whose propagation over the spatial distance calculated by VMC is indeed much
improved and in fact becomes comparable to the coherent Bloch-wave state
characterized by $\mathcal{C}_{ij}^0$ 
as shown in Fig.~\ref{corr}. However, as
indicated in Fig.~\ref{Fig:BlochEnergy}(b), the hole wave function $\varphi_{%
\text{h}}(i)$ is no longer a Bloch wave and $\mathcal{D}_{ij} $ must deviate
from $\mathcal{C}_{ij}^0$ in the long-distance as to be shown below.

Thus, a bare hole created by $c_{i\downarrow}$ on the half-filling vacuum is
no longer a stable elementary excitation to form a conventional Bloch wave.
Instead, in the ground state, the doped hole will break down or
fractionalize to become a new \emph{composite} object, $\tilde{c}%
_{i\downarrow}$, which is composed of mutual spin and charge current
patterns previously shown in Figs. \ref{fig:Js} and \ref{fig:Jh},
respectively. The residual bare hole component has a much reduced
propagation amplitude as indicated in Fig.~\ref{corr}. In the following, we
shall further look into the momentum structure of such a single-hole
propagator in order to follow its long-distance behavior.

\subsection{Momentum distribution $n^h({\mathbf{k}})$ and quasiparticle
spectral weight $Z_{\mathbf{k}}$}

There are two basic physical quantities which can characterize the fate of
the bare hole injected into the half-filling ground state. One is the Landau
quasiparticle spectral weight $Z_{\mathbf{k}}$ defined by\cite%
{Lee1997dispersion}

\begin{eqnarray}
Z_{\mathbf{k}}& \equiv &\left |\langle \text{RVB}| c^{\dagger}_{\mathbf{k}%
\downarrow }|\Psi _{\text{G}}\rangle_{\text{1h}}\right|^2  \notag \\
&=& \frac 1 2 \left |\langle \Psi_{\text{Bloch}}(\mathbf{k})|\Psi _{\text{G}%
}\rangle_{\text{1h}}\right|^2 ~,
\end{eqnarray}
which measures the overlap between the Bloch-wave component (cf. Eq.~(\ref{Bloch}%
)) and the ground state ansatz (\ref{gs}), with ${c}_{\mathbf{k}\downarrow
}^{}=(1/\sqrt{N})\sum_i e^{i{\mathbf{k}}\cdot {\mathbf{r}}_i} {c}%
_{i\downarrow }^{}$. Note that in obtaining the second line on the rhs, $%
\langle {c}^{\dagger}_{\mathbf{k}\sigma}{c}^{}_{\mathbf{k}\sigma}\rangle=1/2$
(i.e., $n^e({\mathbf{k}})=1$) at half-filling is used due to the no double
occupancy constraint.

The second quantity is the hole momentum distribution $n^h({\mathbf{k}})$
defined in Eq.~(\ref{nh}), which is the Fourier transformation of the
single-hole propagator (cf. Eq.~(\ref{C})) 
\begin{equation}  \label{nh1}
n^h(\mathbf{k})=-1+\frac 1 N \sum_{ij \sigma} e^{-i{\mathbf{k}}\cdot \left({%
\mathbf{r}}_i-{\mathbf{r}}_j\right)} \left\langle{c}^{}_{j\sigma }{c}%
^{\dagger}_{i\sigma} \right\rangle_{\text{1h}} ~.
\end{equation}
Here $n^h(\mathbf{k})$ measures the momentum distribution of the hole,
satisfying the sum rule 
\begin{equation}  \label{sumrule}
\sum_{\mathbf{k}} n^h\left( \mathbf{k}\right)=1
\end{equation}
in the single-hole-doped case. Note that $n^e(\mathbf{k})=1$ or $n^h(\mathbf{%
k})=0$ at half-filling for \emph{any} states including excited ones due to
the no double occupancy constraint. So neutral spin excitations (spin
currents) cannot be directly detected by such a momentum distribution.
Nevertheless, beyond $Z_{\mathbf{k}}$, $n^h(\mathbf{k})$ can further show
the momentum change of the bare hole due to the momentum transfer in the
presence of the neutral spin current.

Figures~\ref{t-Jfour}(a) and (c) illustrate the hole momentum distribution $%
n^h\left( \mathbf{k}\right) $ along the cuts of $k_y=\pm \pi/2$ at finite
sizes, and the corresponding $Z_{\mathbf{k}}$'s are presented in Figs.~\ref%
{t-Jfour}(b) and (d). There are totally four peaks located at $\mathbf{K}%
^{0}=\left( \pm \frac{\pi }{2},\pm \frac{\pi }{2}\right)$ as revealed by the
calculated $Z_{\mathbf{k}}$ (cf. Fig.~\ref{Fig1}) in the ground state ansatz
of Eq.~(\ref{gs}). In particular, both $n^h\left( \mathbf{k}\right) $ and $%
Z_{\mathbf{k}}$ in Figs. \ref{t-Jfour}(a) and (b) calculated by VMC based on
Eq.~(\ref{gs}) are in excellent agreement with the DMRG results at the same
sample size of $8\times 8$.

Furthermore, the finite size scalings of $n^h\left( \mathbf{k}\right) $ and $%
Z_{\mathbf{k}}$ of the VMC calculation are presented in Figs. \ref{Fig2}(a)
and \ref{Fig2}(b), where a two-component structure in the ground
state $|\Psi _{\text{G}}\rangle$ in Eq.~(\ref{gs}) is manifested. One is the Bloch-wave
component $|\Psi_{\text{Bloch}}(\mathbf{k})\rangle$ at the momenta $\mathbf{K%
}^{0}$, which gives rise to four peaks in $n^h(\mathbf{k})$ each with the
weight $Z_{\mathbf{K}^{0}}$. However, the weight $Z_{\mathbf{K}^{0}}$
vanishes in a power law fashion (cf. Eq.~(\ref{Zk})). The other is a broad
distribution of the momentum with a weight $n^h(\mathbf{k})\propto 1/N$,
which makes a finite contribution to the sum rule in Eq.~(\ref{sumrule}).
This is consistent with the spin and charge currents presented in the ground
state as shown in Figs.~\ref{fig:Js} and \ref{fig:Jh}, in which the hole and
the background spins share the total momentum such that the bare hole indeed
behaves like an incoherent object with a broad momentum structure.

\subsection{The $\protect\sigma\cdot t\text{-}J$ model: Bloch-wave-like
ground state}

\label{tj-stj}

\begin{figure}[tb]
\begin{center}
\includegraphics[width=0.5\textwidth]{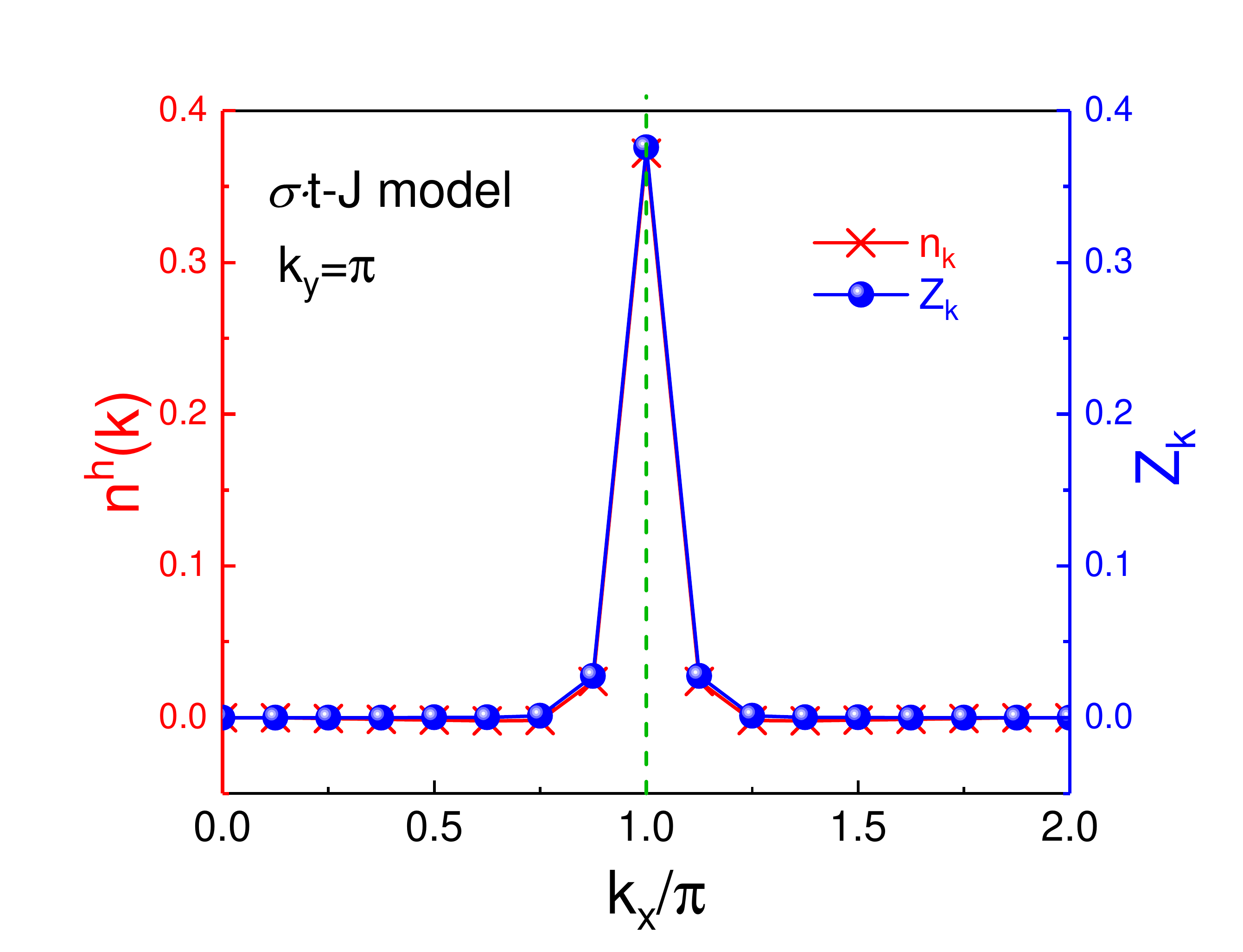}
\end{center}
\par
\renewcommand{\figurename}{Fig.}
\caption{(Color online) Hole momentum distribution $n^h\left( \mathbf{k}%
\right) $ and quasiparticle weight $Z_{\mathbf{k}}$ for the $\protect\sigma%
\cdot t$-$J$ model. Different from the $t$-$J$ model in Fig. \protect\ref%
{t-Jfour}, $n^h\left( \mathbf{k}\right) $ only has single peak located at
the symmetric point $\mathbf{K}_0=(\protect\pi ,\protect\pi)$ with null spin
currents. }
\label{sigmatJnkzk}
\end{figure}

Different from the $t$-$J$ model, we now consider the so-called $\sigma\cdot
t\text{-}J$ model with a modified hopping term\cite{Zhu2013Strong,
Zhu2014nature} 
\begin{equation}
H_{\sigma \cdot t}=-t\sum_{\left\langle ij\right\rangle \sigma }\sigma
c_{i\sigma }^{\dagger }c_{j\sigma }^{}+\text{H.c.}~,  \label{sigma-t}
\end{equation}%
where a spin-dependent sign $\sigma =\pm 1$ is attached to each hopping
process of electron $c_{i\sigma}$. The superexchange term $H_J$ remains the
same as in Eq.~(\ref{Hamiltonian}). It can be shown that the
single-hole-doped $\sigma\cdot t\text{-}J$ model has the same sign structure
as the half-filled Heisenberg model, i.e., the Marshall sign structure\cite%
{Auerbach2012Interacting}. The doped hole moves in the spin background
without creating additional sign mismatches. Therefore, a natural
variational wave function\cite{Wang2015variational} for the $\sigma\cdot t%
\text{-}J$ model is 
\begin{equation}  \label{eq:wf_stJ}
|\Psi _{\text{G}}\rangle _{\text{1h}}^{\sigma \cdot t\text{-}%
J}=\sum_{i}\varphi _{\text{h}}(i)c_{i\downarrow }|\text{RVB} \rangle~,
\end{equation}%
obtained by setting $\hat{\Pi}_i=1$ in Eq.~(\ref{vwf}). It is nothing but
the Bloch-wave state in Eq. (\ref{Bloch}) obtained in the thermodynamic
limit with the translational symmetry.

By minimizing the total energy with the variational parameter $\varphi _{%
\text{h}}$ in Eq.~(\ref{eq:wf_stJ}), the momentum distribution $n^h (\mathbf{%
k})$ and quasiparticle spectral weight $Z_{\mathbf{k}}$ are calculated as
given in Fig.~\ref{sigmatJnkzk}. Different from the $t$-$J$ model, both the
momentum distribution and quasiparticle weight are now sharply peaked at $%
\mathbf{k}=(\pi, \pi) $ without broadening. The angular momentum $L_z$
vanishes without a novel ground state degeneracy, and there are no more spin
and charge currents in the ground state. In other words, the doped hole
simply reduces to a Landau type quasiparticle specified by a momentum at a
symmetric point in the ground state. All are in good agreement with the ED
and DMRG numerical results\cite{Zheng2018}.

\section{Discussion}

\label{Sec::Disc}

To summarize, we have shown that for a single hole injected into a quantum
spin background $|\text{RVB}\rangle$, its ground state is well captured by
the ansatz wave function given in Eq.~(\ref{gs}). Specifically, such a ground
state possesses a nontrivial angular momentum $L_z=\pm 1$ in 2D, which
results in a novel double degeneracy at a given $S^z=\pm 1/2$.
Correspondingly, hidden chiral spin currents around the hole and, \emph{vice
versa}, the chiral hole currents around the spin $S^z=\pm 1/2$ are
identified by the VMC calculation. Such a single-hole state may be further
decomposed into a two-component structure, with a quasiparticle component
characterized by the spectral weight $Z_{\mathbf{k}}$ peaked at four momenta
of $\left(\pm \pi/2, \pm \pi/2\right)$, while there emerges a broad momentum
distribution due to the presence of the neutral spin current which carries
away a partial momentum. These results are in excellent agreement with the
finite-size DMRG calculation. 

Here the wave function of a single hole in a doped Mott insulator is changed
from a simple Bloch-wave to a composite one with the bare hole accompanied
by a neutral spin backflow of many-body nature, i.e., 
\begin{eqnarray}
\varphi_{\text{h}}(i)\propto e^{i{\mathbf{k}}\cdot {\mathbf{r}}_i}
\rightarrow \varphi _{\text{h}}(i)e^{-i\hat{\Omega}_i}~,
\end{eqnarray}
such that the creation operator of the quasiparticle is changed as follows 
\begin{eqnarray}  \label{twist}
{c}_{\mathbf{k}\sigma}^{} \rightarrow \sum_i\varphi _{\text{h}}(i)\tilde{c}%
^{}_{i\sigma } ~.
\end{eqnarray}
Namely, the new ``twisted'' hole is created by $\tilde{c}_{i\sigma }=e^{-i%
\hat{\Omega}_i}c_{i\sigma } $, in which the phase shift $e^{-i\hat{\Omega}_i}
$ (cf.~Eq.~(\ref{Omega})) is solely responsible for the above novel ground
state properties including the finite angular momentum, chiral spin/charge
currents, and the double ground state degeneracy in a 2D system with the $C_4
$ rotational symmetry. In particular, the translational symmetry is
explicitly broken in the variational ground state with $\varphi_{\text{h}}(i)
$ no longer behaving like a Bloch wave. By contrast, $e^{-i\hat{\Omega}_i}$
disappears in the $\sigma\cdot t\text{-}J$ model (cf.~the variational ground
state in Eq.~(\ref{Bloch})) to restore the translational symmetry, in which
the phase string effect is turned off. Clearly, $e^{-i\hat{\Omega}_i}$ is
originated from the phase string effect of the $t\text{-}J$ model. 

As has been emphasized in the Introduction, the conventional spin polaron
effect in the SCBA scheme leads to a coherent quasiparticle picture with a
finite spectral weight $Z_{\mathbf{k}}$ and a narrow but finite bandwidth
for the hole. In a Landau quasiparticle description, such a \emph{rigid}
polaron effect is expected to mainly renormalize the effective mass without
leading to the novel properties discussed in the present approach and the
corresponding ground state wave function is fundamentally distinct without
the persistent neutral spin currents. In general, such a ``longitudinal''
spin polaron\cite{Shraiman1988mobile,
Shraiman1990mobile,Schmitt-Rink1988Spectral, Kane1989,
Martinez1991spin,Liu1991Spectral} or ``spin bag'' effect\cite%
{Weng1988d,Schrieffer1988Spin} may be further incorporated into the present
wave function via $\hat{\Pi}_i$ in Eq.~(\ref{eq:psiG}), which is however
beyond the scope of this work. 

For the present single-hole case, any thermodynamic measurement cannot be directly applied and numerical “experiments” have thus become very useful as employed in the present work. Nevertheless, the novel experimental implications of the present work are indeed very important even though the ground state may be hard to be probed by the spectroscopic measurements. As pointed above, the quasiparticle picture as predicted by the SCBA has been shown to be failed as the doped hole acquires a composite structure. Consequently, it implies that in order to reconcile the well-known discrepancy between the ARPES experiment and the SCBA approach (Refs. 25-28), one should not just try to include the next-neighbor hoppings to improve the dispersion (Refs. 21-24). Rather the line-shape of broadness of the “quasiparticle peak” and its isotropic dispersion observed by the ARPES should be considered together as a reflection of the composite structure or fractionalization of the injected hole (Ref. 44). In particular, the “waterfall phenomenon” at high energy (Ref. 44) should be also understood in the framework of the fractionalization. 

We may generalize the present wave function construction to more hole cases.
For example, the ground state for two holes can be naturally constructed as
follows 
\begin{equation}  \label{pair}
|\Psi _{\text{G}}\rangle _{\text{2h}}=\sum_{ij}g_{ij}\tilde{c}_{i\downarrow }%
\tilde{c}_{j\uparrow }|\text{RVB}\rangle ~,
\end{equation}%
which involves the pairing of two twisted holes instead of bare holes with
an amplitude $g_{ij}$. Indeed, recent VMC calculation\cite{Chen2018two} for
two holes in a two-leg $t$-$J$ ladder has confirmed that by forming such a
bound pair, two holes can significantly gain the kinetic energy by
effectively cancelling out the frustration induced by the phase strings.
There, the variational wave function (\ref{pair}) has been shown to give rise
to the pair-pair correlations in excellent agreement with the DMRG result%
\cite{Zhu2017pairing}. For the $N_h$ case, these twisted holes are expected
to pair up in the ground state of the following form: 
\begin{equation}
|\Psi _{\text{G}}\rangle _{N_\text{h}}=\left( \sum_{ij}g_{ij}\tilde{c}%
_{i\downarrow }\tilde{c}_{j\uparrow }\right) ^{N_h/2}|\text{RVB} \rangle~,
\end{equation}
where the no-double-occupancy constraint is automatically realized in a
half-filling vacuum $|\text{RVB}\rangle$ strictly enforcing the single
occupancy. According to the original RVB theory \cite{Anderson1987Resonating, Anderson1987Resonatingvalence, Zhang1988renormalised}, the binding
potential between holes is originated from the background RVB spin pairing,
but here we emphasize that the emergent phase string effect in $\tilde{c}%
_{\uparrow}$ and $\tilde{c}_{\downarrow}$ will lead to an additional new
pairing force \cite{Chen2018two,Zhu2017pairing} which is nonlocal and dominates over the RVB pairing. Such a
finite doping state has been investigated by a generalized mean-field theory%
\cite{Weng2011superconducting, Ma2014Low} and should be further explored by
VMC in future.

\section{Acknowledgements}

Useful discussions with W. Zheng, Z. Zhu, J. Ho, J. Zaanen, C. Varma are acknowledged. This work is partially supported by Natural Science Foundation of China (Grant No. 11534007), MOST of China (Grant Nos. 2015CB921000 and 2017YFA0302902).
Work by  DNS  was  supported  by  the  Department  of  Energy,  Office of Basic Energy Sciences, Division of Materials Sciences
and Engineering, under Contract No.  DE-AC02-76SF00515
through  SLAC  National  Accelerator  Laboratory.

\appendix
\newpage \onecolumngrid

\section{Variational ground state at half filling} \label{App: GSHF}

At half filling, the $t$-$J$ model is reduced to the spin-$1/2$ Heisenberg
model. Anderson proposed that the ground state should be a \textquotedblleft
resonating valence bond" (RVB) state. The main assumption is that quantum
fluctuations drive the two-dimensional system into a singlet state known as
the spin liquid. This state can be well stimulated by Liang-Doucot-Anderson
type bosonic RVB variational wave function\cite{Liang1988some, Wang2015variational}:%
\begin{equation}
\lvert \text{RVB}\rangle =\sum_{\upsilon }\omega _{\upsilon }|\upsilon
\rangle ,
\end{equation}%
where 
\begin{equation}
|\upsilon \rangle =\sum_{\left\{ \sigma \right\} }\left( \prod_{\left(
i,j\right) \in \upsilon }\epsilon _{\sigma _{i}\sigma _{j}}\right)
c_{1\sigma _{1}}^{\dagger }\cdots c_{N\sigma _{N}}^{\dagger }|0\rangle 
\end{equation}%
is a singlet pairing valence bond state with dimmer covering configuration $%
\upsilon .$ Symbol $\epsilon _{\sigma _{i}\sigma _{j}}$ enables the singlet
pairing between spins on site $i$ and $j.$ The amplitude $\omega _{\upsilon }
$ can be factorized as $\omega _{\upsilon }=\prod_{\left( i,j\right) \in
\upsilon }h_{ij}$ where $h_{ij}$ is a non-negative function depending on
sites $i$ and $j$ of different sublattices. Apparently, such a construction
naturally satisfies the Marshall's sign rule \cite{Wang2015variational} due to the $A$-$B$ sublattice
pairings and the $\epsilon $ tensor. 

\section{Variational Procedure}

The variational procedure involved in this work is essentially the same as presented in Ref. ~\onlinecite{Wang2015variational},  where a single-hole-doped two-leg $t$-$J$ ladder is studied by VMC method based on a ground state ansatz similar to Eq. (\ref{gs}). For the sake of being self-contained,  in the following we outline the main procedures in the VMC calculation and one is referred to Ref.~\onlinecite{Wang2015variational} for more technical details.

\begin{enumerate}
\item The bosonic RVB ground state $|\mathrm{RVB}\rangle $ is optimized (Appendix A) for the superexchange term $H_J$ at half-filling. Upon doping, the ``vacuum state'' $|\mathrm{RVB}\rangle $ is unchanged as the whole change in the spin degrees of freedom as induced by the hole has been attributed to the factor $\Pi _{i}$ in Eq.~(\ref{gs}) generally termed the \emph{spin polaron} effect. 

\item Neglecting the whole spin polaron effect, one has a Bloch-like wave function $|\Psi _{\mathrm{Bloch}}(\mathbf k)\rangle _{\mathrm{1h}}$ in Eq.~(\ref{Bloch}) with
momentum $\mathbf{k}=\left( k_{x},k_{y}\right) $. The corresponding hopping term or the kinetic energy ($E_t\equiv \langle H_t\rangle $) is easily obtained by 
\begin{equation*}
E_{t}=2t_{\mathrm{Bloch}}^{x}\cos k_{x}+2t_{\mathrm{Bloch}}^{y}\cos k_{y}
\end{equation*}%
with
\begin{eqnarray}
t_{\mathrm{Bloch}}^{x,y} &\equiv &\frac{t}{N}\sum_{k,l}\left\langle \mathrm{RVB}\left |c_{k\downarrow }^{\dag
}\left( \sum_{\left\langle ij\right\rangle \sigma }c_{i\sigma }^{\dag
}c_{j\sigma }^{{}}+\mathrm{h.c}\right) c_{l\downarrow }^{{}}\right |\mathrm{RVB}\right\rangle   \notag
\\
&=&\frac{t}{N}\sum_{\left\langle ij\right\rangle }\frac{1}{4}\left( 1+4\left\langle
\mathrm{RVB}|\mathbf{S}_{i}\cdot \mathbf{S}_{j}|\mathrm{RVB}\right\rangle \right) ~.
\end{eqnarray}%
The numerical simulation based on the VMC shows that $t_{\text{Bloch}%
}^{x,y}<0$ which leads to the minimal energy state $|\Psi _{\text{Bloch}}(\mathbf{k})\rangle _{\mathrm{1h}}$ at momentum $\mathbf{k}=\left( 0,0 \right) $
as presented in Fig. \ref{Fig:BlochEnergy}(a).    

\item Based on the general variational ground state ansatz in Eq. (\ref{gs}), the kinetic energy can be expressed by 
\begin{equation}
E_{t}=-\sum_{\left\langle ij\right\rangle }\left( \tilde{H}_{t}\right)
_{ij}\varphi _{h}^{\ast }\left( j\right) \varphi _{h}\left( i\right) +\mathrm{h.c}
\end{equation}%
where $\widetilde{H}_{t}$ is given by
\begin{equation}\label{tildeHt}
\left( \widetilde{H}_{t}\right) _{ij}\equiv -t\sum_{\sigma }\left\langle
\mathrm{RVB}|c_{j\downarrow }^{\dag }c^{}_{j\sigma }e^{-i\left( \hat{\Omega}_{j}-\hat{\Omega}%
_{i}\right) }c_{i\sigma }^{\dag }c^{}_{i\downarrow
}|\mathrm{RVB}\right\rangle ~,
\end{equation}
which can be calculated \cite{Wang2015variational} directly by the VMC.  Similarly, we can obtain the superexchange energy $E_J\equiv \langle H_J\rangle$,
\begin{equation} \label{tildeHJ}
E_J=\sum_{i} (\widetilde{H}_J)_{i} |\varphi_h(i)|^2
\end{equation}
with 
\begin{equation}
(\widetilde{H}_J)_{i}=J\sum_{\substack{ \left\langle kl\right\rangle  \\ k\neq i,l\neq i}}\left\langle \mathrm{RVB}\left| e^{i\hat{\Omega}_i}\mathbf{S}_k\cdot\mathbf{S}_l e^{-i\hat{\Omega}_i}n_{i}^{\downarrow}\right |\mathrm{RVB}\right\rangle~.
\end{equation}
In principle, the ground state wave function $\varphi_{h}\left( i\right)$ can be determined by diagonalizing a single-particle effective Hamiltonian $%
\hat{H}_{\mathrm{eff}}\equiv \widetilde{H}_{t}+\widetilde{H}_J$. For a large-size square lattice, the superexchange matrix element $(\widetilde{H}_J)_i$ has essentially the same value for different hole positions in the bulk due to translational symmetry. Thus, the term $\widetilde{H}_J$ plays a negligible role in determining $\varphi_{h}\left( i\right)$. Instead, the wave function $\varphi_{h}\left( i\right)$ can be optimized by directly diagonalizing  $\widetilde{H}_t$ to result in $E_t$ with a constant $E_J$.
\end{enumerate}

%

\end{document}